\documentclass[pre,preprint,showpacs]{revtex4-1}

\usepackage{graphicx}

\usepackage{graphics}
\usepackage{amsmath}
\usepackage{psfrag}
\usepackage{graphicx}
\usepackage{amsthm}

\theoremstyle{definition}

\begin{document}

\title{Ensemble Inequivalence and Maxwell Construction in the Self-Gravitating Ring Model}

\author{T.~M.~Rocha Filho}
\email{marciano@fis.unb.br}
\affiliation{Instituto de F\'{i}sica and International Center for Condensed Matter Physics,
Universidade de Bras\'\i lia, 70910-900 - Bras\'\i lia, Brazil}
\author{C.~H.~Silvestre}
\affiliation{Instituto de F\'{i}sica, Universidade de Bras\'\i lia, 70910-900 - Bras\'\i lia, Brazil}
\author{M.~A.~Amato}
\affiliation{Instituto de F\'{i}sica and International Center for Condensed Matter Physics,
Universidade de Bras\'\i lia, 70910-900 - Bras\'\i lia, Brazil}

\begin{abstract}

The statement that Gibbs equilibrium ensembles are equivalent is a base line in many approaches in the context of
equilibrium statistical mechanics. However, as a known fact, for some physical systems this equivalence may not be true. In this
paper we illustrate from first principles the inequivalence between the canonical and microcanonical ensembles
for a system with long range interactions. We make use of molecular dynamics simulations and Monte Carlo
simulations to explore the thermodynamics properties of the self gravitating ring model and discuss on
what conditions the Maxwell construction is applicable.
\end{abstract}


\maketitle

\section{Introduction}

Equilibrium Statistical Mechanics is a hallmark of theoretical physics and an invaluable tool to study the properties of
matter for more than a century. At its foundations lies Gibbs' ensembles theory~\cite{gibbs}, which is an elegant formulation
applicable to a very broad class of phenomena (for a brief history of ensemble theory see~\cite{inaba}).
As it is well known for any rounded practitioner, the microcanonical ensemble is harder to use
than the canonical ensemble, and the grand-canonical ensemble is in a sense the more simple among them. Provided that
the predictions of different ensembles coincide, one can then choose which one to use according to the needs in consideration.
As a consequence, many authors devoted a considerable effort to the task of proving and establishing the limits of validity of the
equivalence of the different ensembles (see ~\cite{lewis} and references therein). From the thermodynamic viewpoint
ensemble equivalence is based on the fact that the Legendre transformation connecting different ensembles is invertible.
This implies particularly that the entropy is a concave function of the energy in the whole physically accessible energy range.
In statistical terms it means that all properties of the system are well described either in terms of energy or temperature,
that is essentially Gibbs' argument.

These respective ensembles are equivalent in the thermodynamic limit $N\rightarrow\infty$ if the interaction potential is tempered and stable, i.~e.\
if the energy is additive and a stable equilibrium state exists~\cite{ruelle}.
The Helmholtz free energy and the grand-potential are then obtained from the microcanonical entropy by the usual Legendre transforms.
Examples of physically relevant and non-stable potentials are the gravitational interaction, where for some specific cases the non-stability leads to the
so-called gravothermal catastrophe~\cite{binney}, and multi-species plasmas~\cite{ruelle}.
As well known, equilibrium ensembles for self-gravitating systems are inequivalent, and an energy interval with negative heat capacity
exists in the microcanonical ensemble~\cite{padmanabhan}. The appearance of a convexity
region in the entropy-energy curve breaks down the equivalence and
any state of the system in the convex region cannot be realized in the canonical ensemble.
The simplicity of Gibbs argument fails in this case~\cite{ellis}.

According to van Hove's theorem~\cite{hove}, and under certain assumptions, the pressure $P$ in the canonical ensemble
must be a decreasing function of the volume $V$ and becomes constant for the values of $V$
in the interval of phase coexistence. In the microcanonical ensemble this corresponds to the Maxwell construction prescription to replace
the entropy by its concave envelope.
One of the conditions required in van Hove's theorem is that, for
a three-dimensional system, the interparticle potential $V(r)$ satisfies $V(r)\geq r^{-3-\alpha}$ for $\alpha>0$ and large distances $r$,
i.~e.\ that the potential is short-ranged and the total energy is additive.
Therefore, in equilibrium statistical mechanics calculations, a convex intruder
in the entropy function can only exist as a result of approximations,
e.~g.\ using a mean-field approach for a system with short-range interactions,
or as a consequence of finite size effects~\cite{wales,lyndenbell,gross}. This point is very well illustrated
for the two-dimensional Potts model with nearest neighbors interaction, where a convex dip is present for small lattice size,
disappearing for increasing $N$ with the negative specific heat region being replaced by a flat curve,
while for globally coupled spins, the convex dip remains
even in the thermodynamic limit~\cite{ispolatov}.

Thus the additivity of energy and entropy, which follows from the temperedness of the potential,
and its stability, ensure that equilibrium ensembles are equivalent. The important point is that there exist real systems
for which these conditions are not met. However, this does not imply that ensembles are not equivalent as the conditions are sufficient but not necessary. 

Besides self-gravitating systems, examples of real physical situations with the occurrence of a convex intruder in the entropy are
two-dimensional quasi-geostrophic flows~\cite{turkington}, wave-particle interaction in a plasma in the presence
of two harmonics~\cite{elskens,teles} and magnetically self-confined plasma torus~\cite{kiessling}.
They are also
are examples of long-range interacting systems, with interacting potentials decaying at long-distances
as $1/r^\alpha$ with $\alpha<D$, $D$ being the spatial dimension~\cite{proc1,proc2,proc3,booklri,campa,levin}.
This definition implies a non-additive energy as the interaction energy between two subsystems is no longer
negligible when compared to the bulk energy.
It is worth noticing that this definition may be at variance with some works in the literature, as for instance in Ref.~\cite{kac}
where an interaction with an exponential dependence on distance, and therefore not long-ranged in the sense adopted here, is referred as long-ranged.

The main goal of the present paper is to illustrate with a specific model of a many particle system with dynamics,
for the first time up to the author's knowledge, the inequivalence of the microcanonical and canonical ensembles
from first principles, i.~e.\ by only solving the Hamilton equations of motion, and to show
that the results so obtained are in agreement with previous theoretical studies and Monte-Carlos simulations.
This allows us to discuss the physical origin of ensembles inequivalence
as well as the meaning of the Maxwell construction if the interactions are long-ranged.
The model system chosen is the one-dimensional self-gravitating ring model with Hamiltonian in Eq.~(\ref{ringmodham}), as its
thermodynamic properties are well known, with a first order phase transition from a homogeneous to a non-homogeneous phase~\cite{sota}.

The structure of the paper is the following: the ring model is presented in Section~\ref{ringmod} and
in Section~\ref{mce} we present our molecular dynamics results and compare them to Monte Carlo simulations and results
from previous works. In Section~\ref{disc} we discuss our results and present some concluding remarks.

\section{The self-gravitating ring model}
\label{ringmod}

The Self-Gravitating Ring (SGR) model was introduced by Sota and collaborators~\cite{sota}
and describes a system of $N$ particles constrained to move on a circle and interacting by
a gravitational potential regularized by a (usually small) softening parameter $\epsilon$ introduced
in order to avoid the divergence of the potential at short distances.
With a proper choice of units its Hamiltonian can be written as:
\begin{equation}
H=\sum_{i=1}^N\frac{p_i^2}{2}-\frac{1}{2N}\sum_{i,j=1}^N\frac{2\sqrt{\epsilon}}{\sqrt{1-\cos(\theta_i-\theta_j)+\epsilon}},
\label{ringmodham}
\end{equation}
with $\theta_i$ being the position angle on the circle of particle $i$ and $p_i$ its conjugate (angular) momentum. Here units have been
chosen such that the minimum value of the energy per particle is $-1$ irrespective of the value of $\epsilon$. The factor $1/N$  in the potential energy
term is known as the Kac factor, and can be introduced by a change of time units (as long as $N$ remains finite), in order for the total
energy to be extensive, although remaining non-additive. It also facilitates the comparison of results with different numbers of particles.
The analogous of a magnetization can be introduced here by its components:
\begin{equation}
M_x=\frac{1}{N}\sum_{i=1}^N\cos\theta_i,\hspace{10mm} M_y=\frac{1}{N}\sum_{i=1}^N\sin\theta_i.
\label{magscompsdef}
\end{equation}
Many properties of the model were studied in previous works~\cite{eu1,casetti,monechi,tatekawa,nardini,eu2}. It has a phase-transition
from a low energy ferromagnetic phase to a high energy homogeneous (non-magnetic) phase. The order of the transition depends on the value of
the softening parameter. For smaller values of $\epsilon$ the transition is first order and becomes continuous for higher values
of the parameter. It is worth noticing that for systems with long-range interaction in the $N\rightarrow\infty$ limit
particles are exactly uncorrelated~\cite{chavanis}.

\section{Microcanonical and canonical ensembles for the self-gravitating ring model}
\label{mce}

\begin{figure}[ptb]
\begin{center}
\scalebox{0.3}{\includegraphics{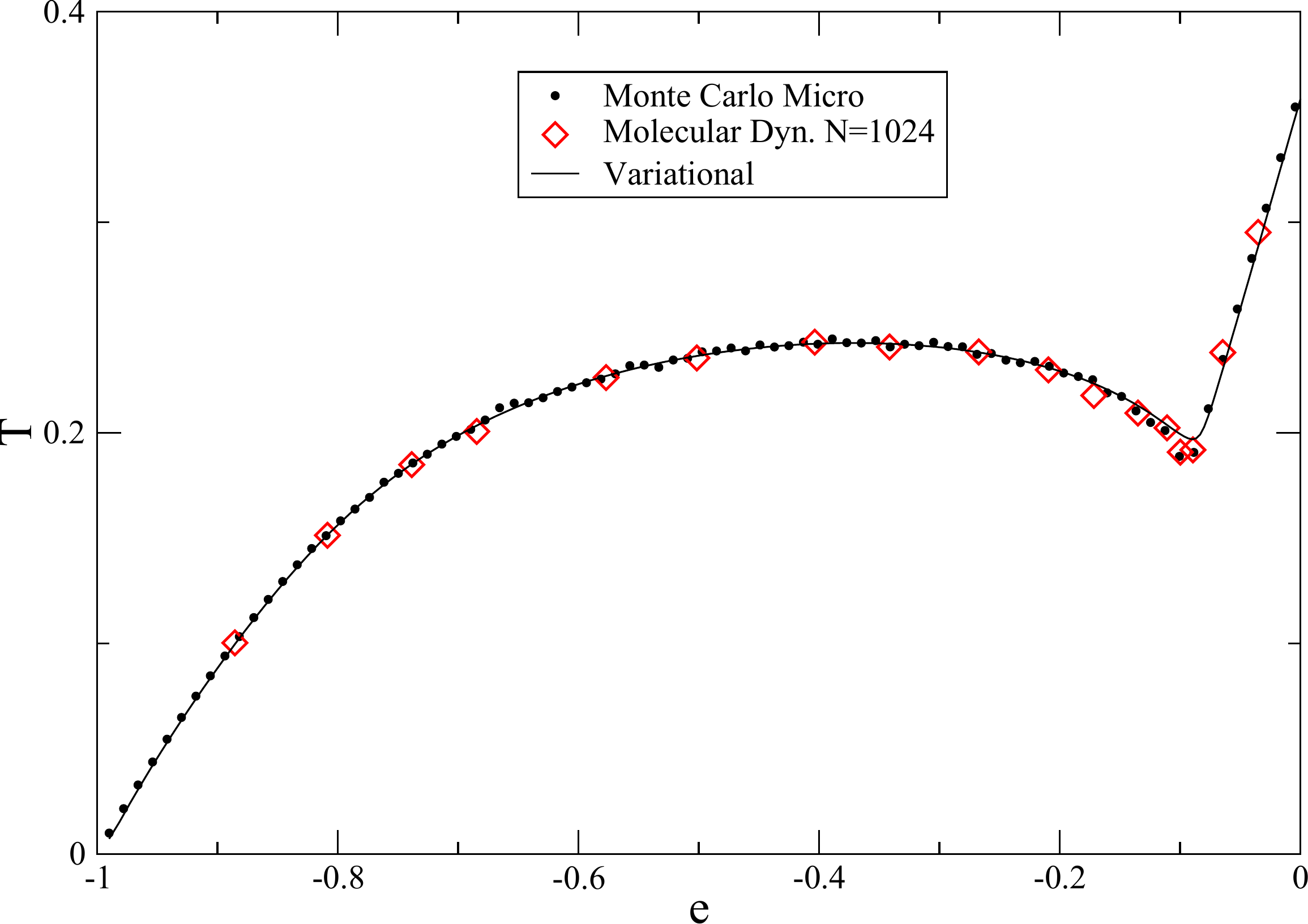}}
\end{center}
\caption{Caloric curve for the ring model with $\epsilon=10^{-2}$ from the variational method,
microcanonical Monte Carlo simulations with $N=1000$, and microcanonical molecular dynamics simulation with $N=1024$ using a time
step  $\Delta t=0.02$.}
\label{caloricmicrofig}
\end{figure}

We fix the value for the softening parameter as $\epsilon=10^{-2}$ such that the system has a continuous phase transition
and a negative heat capacity for an energy interval, at the same time allowing for faster molecular dynamics simulations
(smaller values of $\epsilon$ leads to higher values of numeric error at fixed time step in the integration algorithm).
For reference purposes we first determine the caloric curve for the system for the chosen value of $\epsilon$, with very high accuracy,
using an iterative numeric variational method by Tatekawa et al.\ that also applied it for studying the thermodynamics of the SGR model~\cite{tatekawa}.
It amount to maximizing the Gibbs entropy for uncorrelated particles:
\begin{equation}
s\equiv S/N=-\int{\rm d}p\:{\rm d}\theta\:f_1(p,\theta)\ln f_1(p,\theta),
\label{gibbsent}
\end{equation}
where the Boltzmann constant is set to unity and $f_1(p,\theta)$ is the one-particle distribution function.
The results are shown in Fig.~\ref{caloricmicrofig}, where a region of negative heat capacity is clearly visible.
Note that the variational method is solely
based on entropy maximization on the space of one-particle distribution functions, and the results obtained
are thus a direct consequence of the second law of thermodynamics.
Microcanonical Monte Carlo results are also shown in the figure and were obtained using the method described in Ref.~\cite{ray},
with a very good agreement with the variational method.

\begin{figure}[ptb]
\begin{center}
\scalebox{0.3}{{\includegraphics{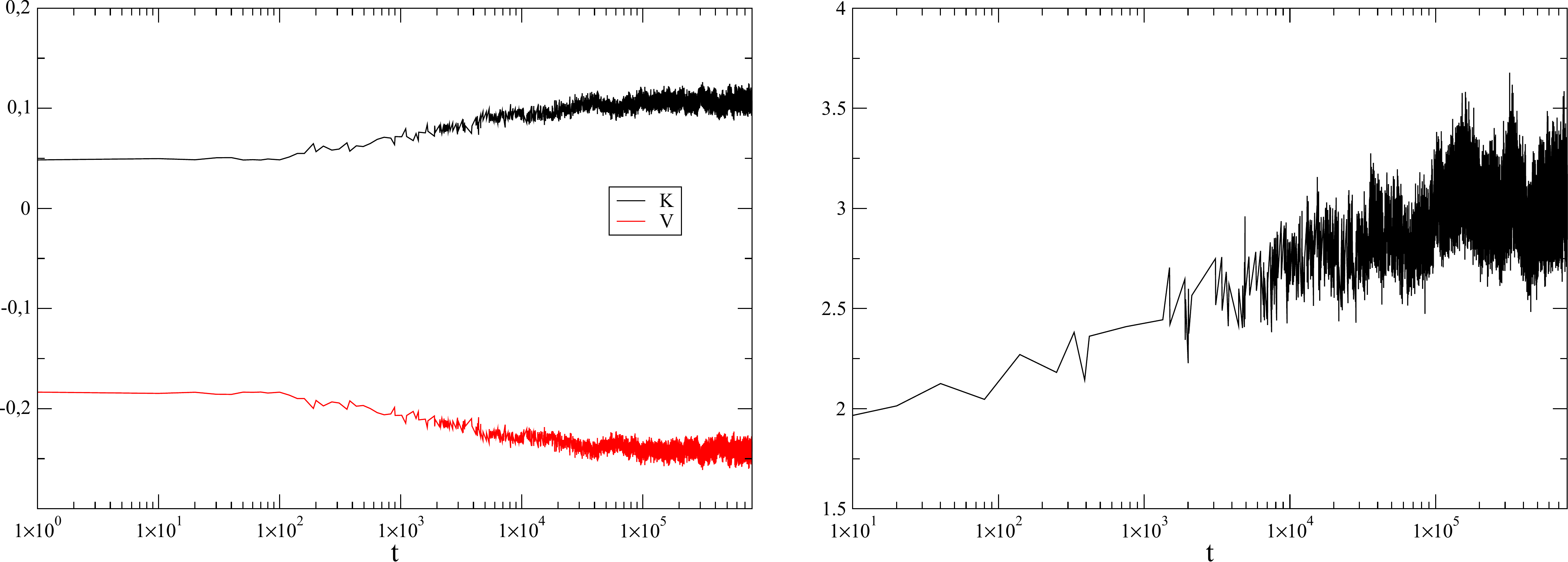}}}
\end{center}
\caption{Left Panel: Kinetic $K$ and potential $V$ energies per particle as a function of time for a homogeneous waterbag initial state
with total energy per particle $e=-0.135$ which correspond to an equilibrium state with a negative heat capacity,
simulation time step $\Delta t=0.02$ with a relative error in the energy of order $10^{-6}$.
Right Panel: Kurtosis as a function of time.}
\label{exdynmicro}
\end{figure}

One may ask whether the system can relax to an equilibrium state if its energy belongs to the region with a negative heat capacity.
In order to answer that question we numerically solve the Hamiltonian equations of motion for the $N$ particle system
using a graphics processing unit parallel implementation~\cite{eu3}
of a fourth-order symplectic integrator~\cite{yoshida}. The system is initially prepared in a {\it waterbag} non-equilibrium state
with a uniform distribution in the intervals $-p_0<p<p_0$ and $-\theta_0<\theta<\theta_0$, with the constants $p_0$ and $\theta_0$
chosen for the system to have the required energy. Since the relaxation time to reach equilibrium is typically very long
in long-range interacting systems, and scales with
$N$ for non-homogeneous or $N^2$ for homogeneous states~\cite{eu2,eu4,chris}, very long computer runs are required.
The left panel of Fig.~\ref{exdynmicro} shows the time evolution of the total kinetic $K$ and potential $V$ energies per particle for
a homogeneous waterbag non-equilibrium initial state with $\theta_0=\pi$ and total energy per particle $e=-0.135$.
The total simulation time for this case is roughly 12 hours on a NVIDIA GTX 690 graphic card.
In order to measure the distance to equilibrium we use the kurtosis of the momentum distribution, i.~e.\ the fourth reduced statistical moment of $p$ given by
${\cal K}=\langle p^4\rangle/\langle p^2\rangle^2$. Its value for any Gaussian distribution is given by ${\cal K}=3$.
The right panel of Fig.~\ref{exdynmicro} shows the kurtosis of the momentum distribution, where
we see that the system reaches equilibrium at $t\approx2\times10^5$.
Data points obtained from such simulations and different energy values are also shown in 
Fig.~\ref{caloricmicrofig}, with a very good agreement with both microcanonical Monte Carlo simulations and the variational method.
Thus, at least for the present model, equilibrium states with a negative heat capacity are accessible by relaxation from non-equilibrium states.

\begin{figure}[ptb]
\begin{center}
\scalebox{0.3}{{\includegraphics{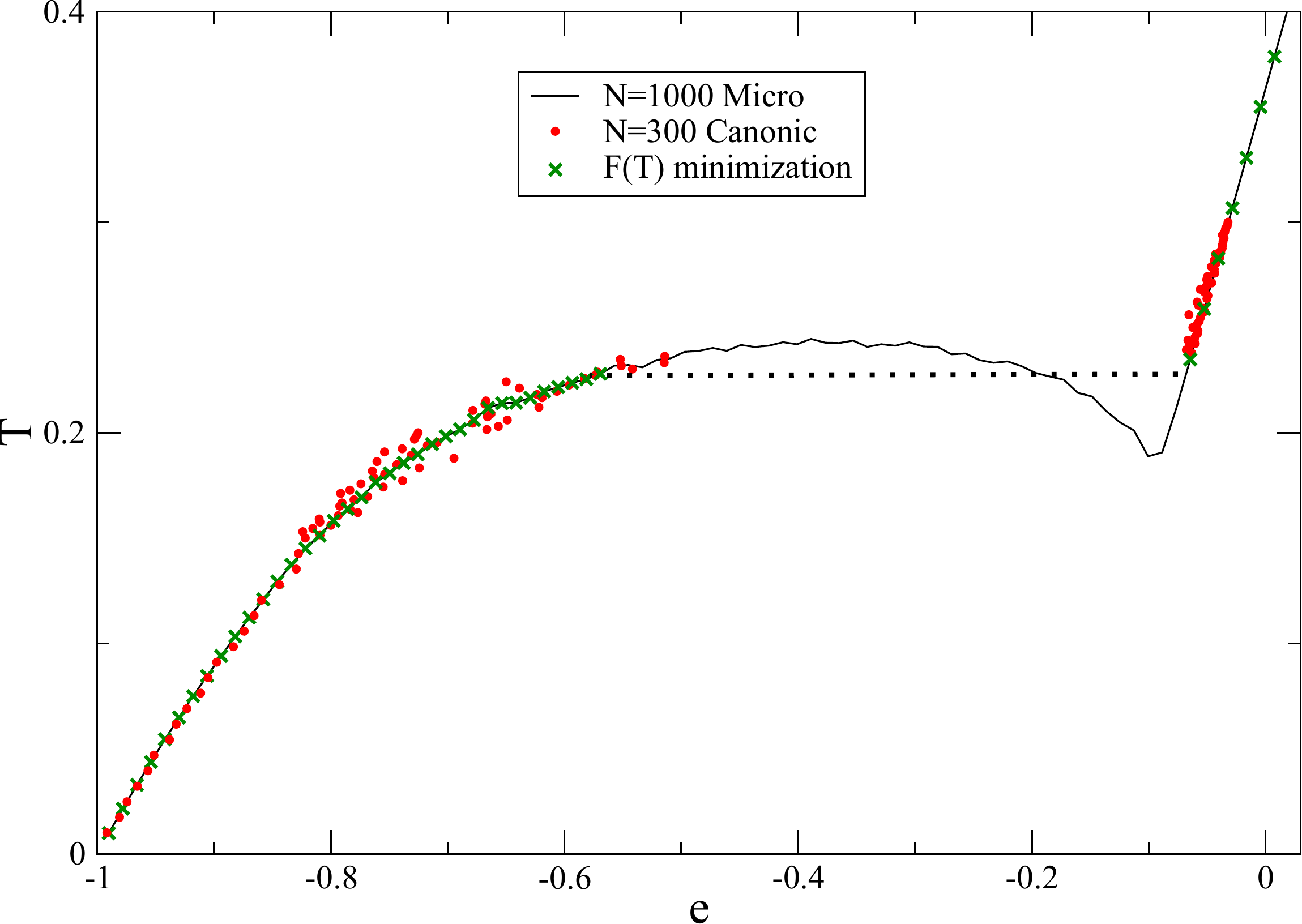}}}
\end{center}
\caption{Caloric curve for the ring model with $\epsilon=10^{-2}$ from microcanonical and canonical Monte Carlo
simulations with $N=1000$ and $N=300$, respectively. The crosses correspond to the minimization of the Helmholtz free energy and
the dotted line the phase coexistence curve predicted by the Maxwell prescription (see text).}
\label{ensemblestufig}
\end{figure}

The next point to consider is to determine the caloric curve in the canonical ensemble. Canonical Monte Carlo results are show in
the left panel of Fig.~\ref{ensemblestufig}, alongside with points obtained from the minimization of the free energy.
The latter was obtained using the usual relation $F=E-TS$ with the entropy computed from Eq.~(\ref{gibbsent}) by writing
\begin{equation}
f_1(p,\theta)=\sqrt{\beta/2\pi}\:e^{-\beta p^2/2}\rho(\theta)
\label{onepartdf}
\end{equation}
and the (normalized) spatial distribution $\rho(\theta)$ given a histogram of the particle positions from
the microcanonical Monte Carlo simulation.
Both results from canonical Monte Carlo and free energy minimization are in quite good agreement except for a few points
corresponding to metastable states close to $e=-0.55$. Such states, being local minima of the free energy,
trap the Monte Carlo evolution for a very large number of steps and are known to cause numerical difficulties.
To circumvent the effects of such metastable states we used a relatively small number of particles $N=300$.

\begin{figure}[ptb]
\begin{center}
\scalebox{0.3}{{\includegraphics{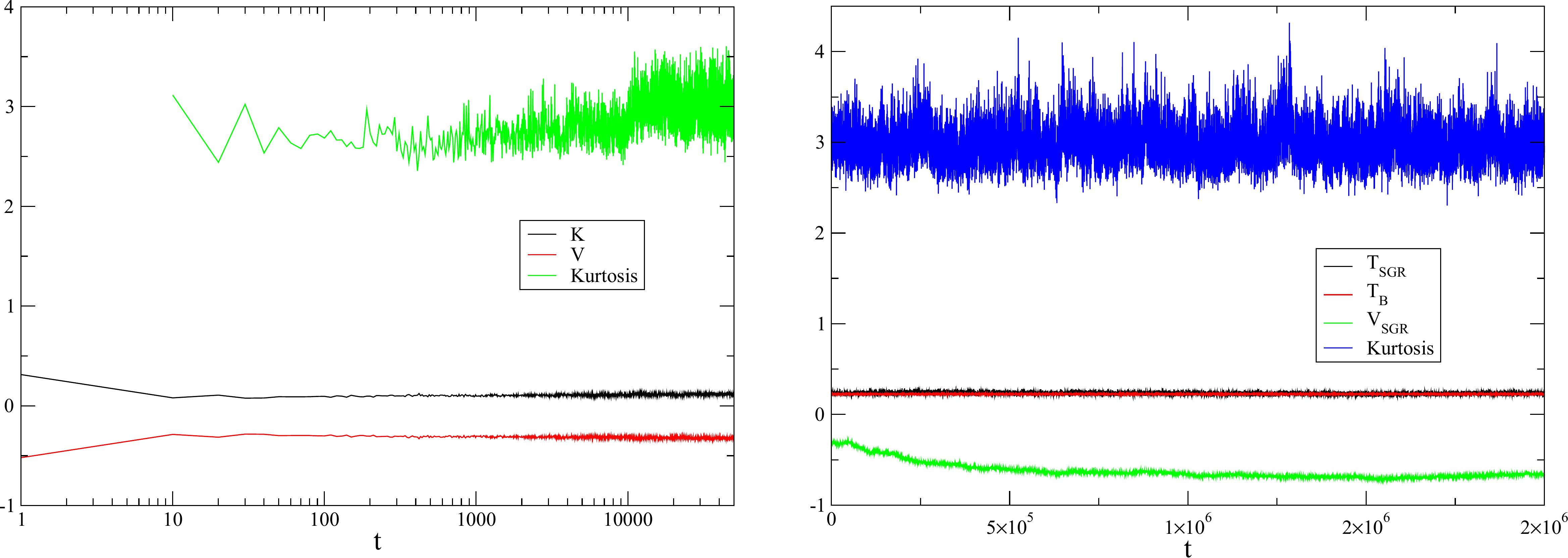}}}
\end{center}
\caption{Left Panel: Evolution to thermodynamic equilibrium of a waterbag initial state with energy per particle $e=-0.2$.
Right Panel: Time evolution of the same microcanonical equilibrium state at energy per particle $e=-0.2$
of the left panel when in contact with a heat bath at the same temperature showing the SGR model ($T_{SGR}$) and bath ($T_B$) temperatures.
the model potential energy ($V_{SGR}$) and the kurtosis of the momentum distribution of the SGR model.
The simulation parameters are $\Delta t=0.02$, $t_f=2\times 10^6$.}
\label{energsfig}
\end{figure}

Now, if the system, in a microcanonical equilibrium state with a negative heat capacity, is put in contact with a thermal bath
at the same temperature, it becomes unstable and evolves to the canonical equilibrium state that minimizes the free energy,
for the same temperature of the initial state.
We illustrate this, again from from first principles, by performing a molecular dynamics simulation of both system and thermal bath.
Following the same approach as in Ref.~\cite{nosprl}, the bath is modeled by a Hamiltonian system with a short-range
interaction formed by $M$ rotors with Hamiltonian:
\begin{equation}
H_{bath}=\sum_{i=N+1}^{N+M}\frac{p_i^2}{2}+\sigma\sum_{i=N+1}^{N+M}\left[1-\cos(\theta_i-\theta_{i+1})\right],
\label{bathham}
\end{equation}
where $\theta_i$ and $p_i$ the coordinates and the momenta of particle $i$, with $\theta_{N+M+1}\equiv\theta_{N+1}$
and $\sigma$ the interaction strength among first neighbors. The number of particles in
the bath is chosen such that $M\gg N$ to guarantee that the main system only causes minor disturbances in the bath.
The SGR model and the bath are coupled by the interaction potential:
\begin{equation}
V_{int}=\lambda\sum_{i=1}^L\left[1-\cos(\theta_i-\theta_{i+N})\right],
\label{intpotsb}
\end{equation}
with $\lambda$ the coupling parameter and $L$ chosen typically as a fraction of $N$.
Since the initial conditions for both systems are randomly chosen from given initial distribution,
any choice for which particles are coupled is equivalent, and we simply chose to couple the $L$ particles
of the self-gravitating ring model with indices $i=1,\dots,L$ to the $L$ particles of the bath
with indices $i=N+1,\dots,N+L$.
The total Hamiltonian is then the sum of the Hamiltonian in Eq.~(\ref{ringmodham}), $H_{bath}$ and $V_{int}$.
The choice of the parameters requires some experimentation,
and the values used here are $N=512$, $M=204\:800$, $L=16$, $\sigma=5.0$ and $\lambda=1.0$.
The left panel in Fig.~\ref{energsfig} shows the time evolution to equilibrium of the isolated SGR model with a waterbag initial state
with energy $e=-0.2$, which lies inside the negative heat capacity energy interval. The resulting microcanonical equilibrium state is then
coupled to the thermal bath. The time evolution of temperature as given by twice the kinetic energy, the kurtosis of the
velocity distribution for the main system, and the interaction energy between the system and the bath are show in the right panel of the
same figure. The ring model starts from the initial state become unstable and evolves to a canonical equilibrium while
the temperature remains constant, up to small fluctuations. The kurtosis remains always close to the equilibrium value,
indicating that the momentum distribution function remains a Gaussian during the whole time evolution. The spatial
distribution function changes until the final state corresponds to the minimum of the free energy for the non-homogeneous phase.
\begin{figure}[ptb]
\begin{center}
\scalebox{0.3}{{\includegraphics{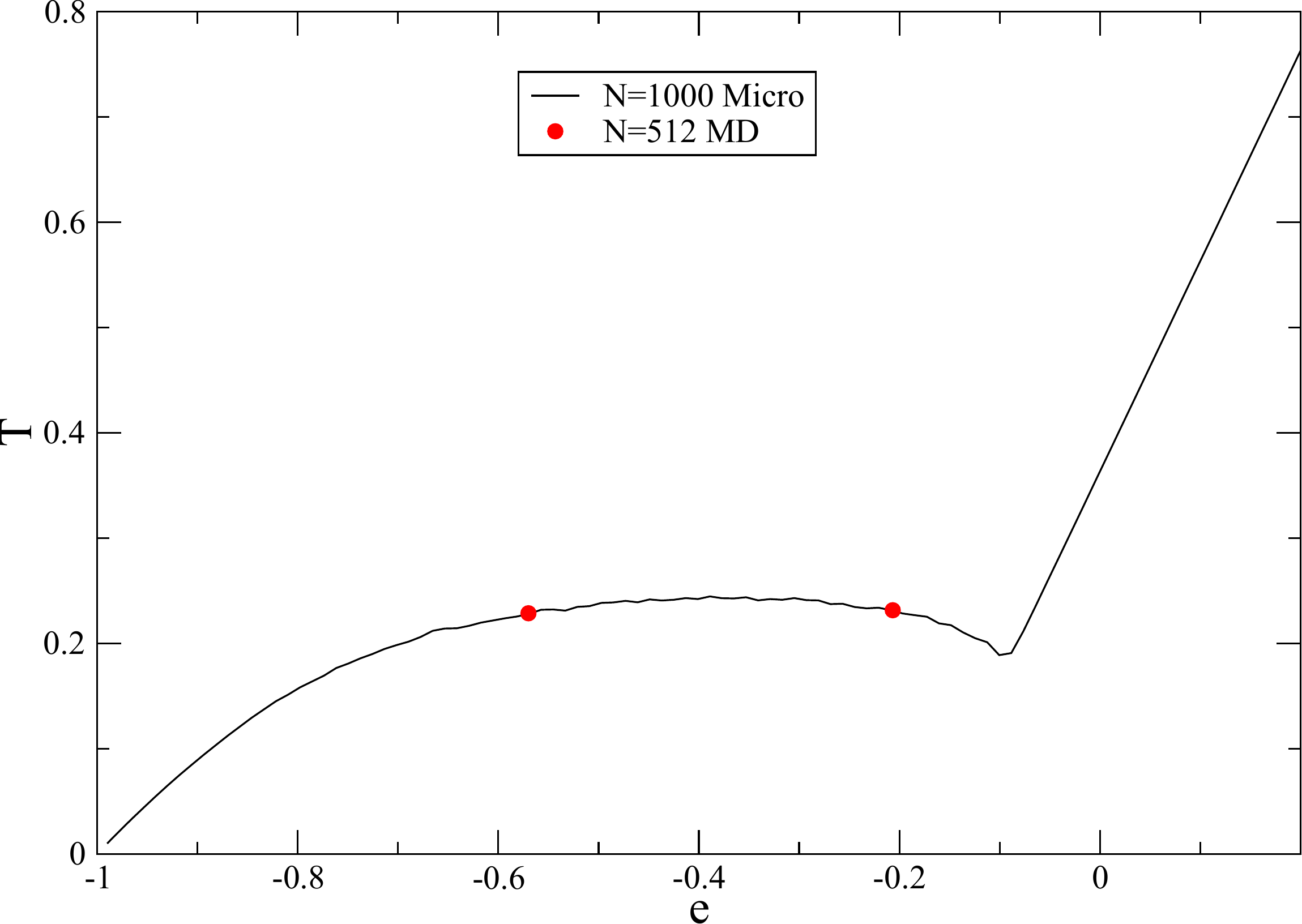}}}
\end{center}
\caption{Initial ($e=-0.2$) and final ($e\approx-0.57$) states for a microcanonical stable state in the negative heat capacity region after turning
on the coupling to a heat bath. the ring model system has $N=512$ particles and the bath $N_b=204\:800$ particles.
The simulation parameters are $\Delta t=0.02$, $t_f=2\times 10^6$.}
\label{bathstatesfig}
\end{figure}

\begin{figure}[ptb]
\begin{center}
\scalebox{0.3}{{\includegraphics{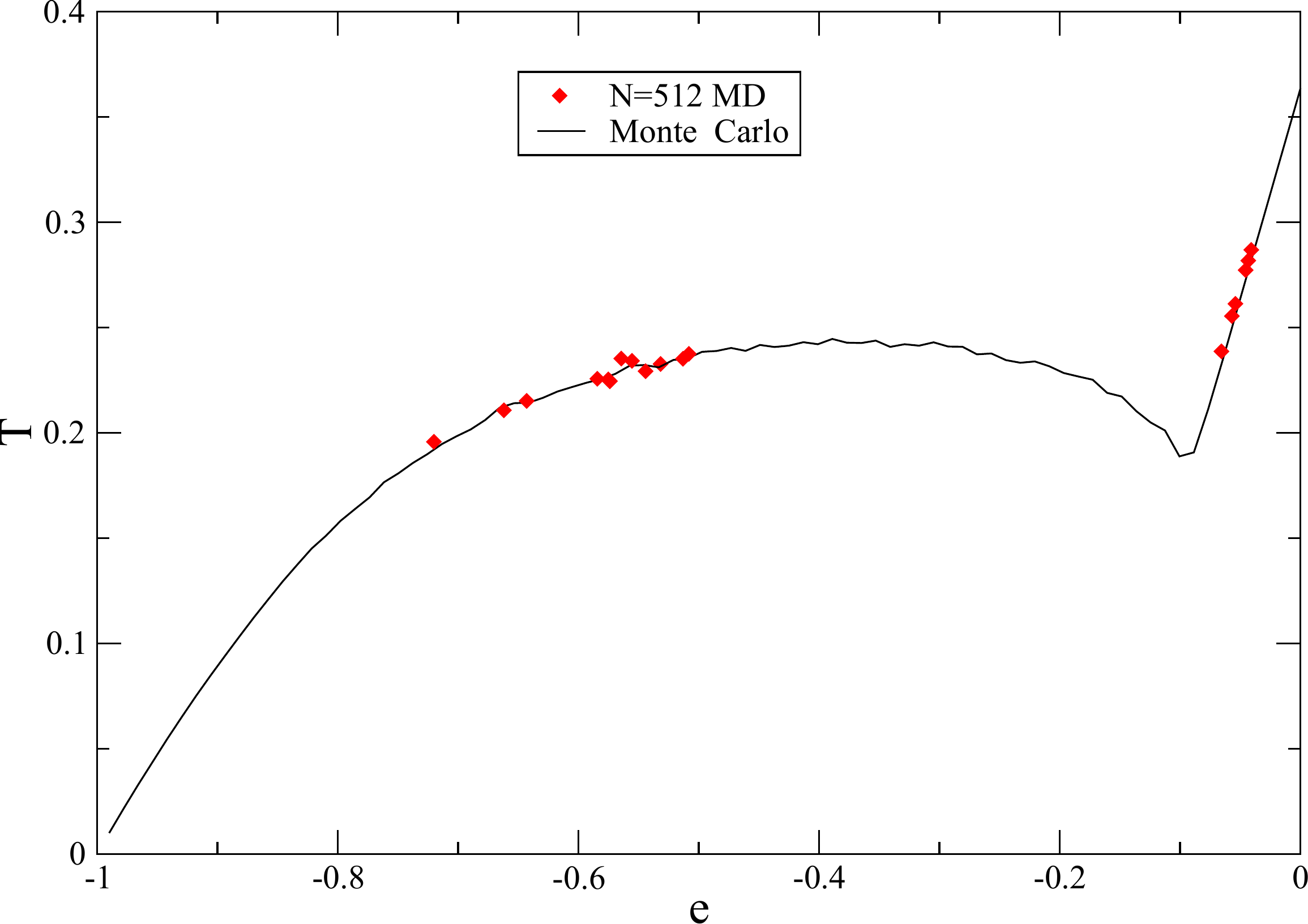}}}
\end{center}
\caption{Temperature as a function of total energy per particle $e$ (diamonds) from Molecular Dynamics (MD) simulation in the canonical ensemble
for $N=512$ particle in the ring-model and $N_b=204\:800$ particles in the bath. The simulation parameters are the
same as in Fig.~\ref{bathstatesfig}. The continuous line is the microcanonical Monte Carlo caloric curve drawn for comparison. Each MD
simulation point in the graph requires roughly $18$ hours of computer time.}
\label{caloriccan}
\end{figure}

The caloric curve in the canonical ensemble from MD simulations is obtained in the following way: 
the system is prepared in a low energy equilibrium state, corresponding to a positive heat capacity in contact with the heat bath.
The temperature of the bath is then raised by a factor $\alpha$ by multiplying the velocities of the particles in the bath by $\sqrt{\alpha}$.
The coupled system and bath are then left to evolve, after a very long time, into a new equilibrium state, and the total energy and temperature
as twice the kinetic energy are obtained from a numeric average over a given time interval.
The results are show in Fig.~\ref{caloriccan}, where the jump in energy is clearly visible, in accordance with canonical Monte Carlo results
shown in Fig.~\ref{ensemblestufig}.
By comparing Figs.~\ref{ensemblestufig} and~\ref{caloriccan} we see that when rising the temperature from a stable non-homogeneous state,
the jump occurs at an energy value dictated by the minimization of the free energy, as obviously expected.
For systems with short-range interactions, states in the gap are accessible by considering different proportion of the particles in each one
of the phases at each extremity of the gap. This leads to the Maxwell construction prescription and yield
the dotted line in Fig.~\ref{ensemblestufig}.
On the other hand, for long-range interacting systems and as extensively discussed in the pertaining literature (see for instance Ref.~\cite{booklri}),
there is no phase-coexistence and the line associated to the Maxwell construction is physically meaningless.
Our numerical experiment for the SGR model shows this behavior clearly as a small increase in the temperature at the phase transition causes
a discontinuous jump in the total energy. This is different to what is observed in the microcanonical ensemble.
We stress here that this comes out directly and solely from the dynamics of the system, i.~e.\ from the numeric solution of the
Hamilton equations of motion.

\section{Discussion and Conclusions}
\label{disc}

Up to the authors' knowledge, the verification for a specific model of ensemble inequivalence was never shown before from full
$N$-body molecular dynamics molecular simulations and without any other assumption on the system.
From the results above we see that for the ring model in the canonical ensemble, i.~e.\ for a system coupled to a large energy reservoir,
the whole energy interval corresponding to the Maxwell construction is not physically realizable in the canonical ensemble.
If the temperature is raised starting from a non-homogeneous state with positive heat capacity, the energy jumps discontinuously,
the energy interval of the jump corresponding to the usual Maxwell prescription, but without phase coexistence that would
correspond to this inaccessible energies in the canonical ensemble.
If the system is in a state with energy in this interval, it will absorb from or give energy to the
bath until it reaches a stable canonical equilibrium that minimizes the free energy, as expected~\cite{touchette}, and as illustrated
from our direct numerical simulations. This is at variance to systems with short range interactions where any energy in the interval 
where the Maxwell construction is required is realizable by a combination of the two different phases, as guaranteed by the additivity
of energy and entropy valid for such systems but not for long-range interactions.

It is a common practice, when computing statistical mechanics properties of a short-range interacting system, to rely on some approximation that results
in a van der Waals loop, e.~g.\ using a mean-field approximation. In this case replacing the convex dip in the entropy function
by a flat line according to the Maxwell construction prescription is fully justified, as the entropy of the original system is additive,
although this property is usually not satisfied in a mean-field approximation. On the other hand, if the system has long-range interactions,
or equivalently is not tempered and does not satisfies the condition of van Hove's theorem, then the convex dip may not disappear
in the thermodynamic limit. Using the Maxwell construction in this case is plainly wrong as it leads to physically non
realizable states. In this case, the correct prescription is to take the
whole region corresponding to a negative heat capacity as non physical.
A purely thermodynamic description of long-range interacting systems remains possible along the lines discussed
in Refs.~\cite{latella1,latella2}.

\section{Acknowledgments}

The authors would like to thank S.~R.~Salinas and V.~B.~Henriques for fruitful discussions. TMRF was partially financed by CNPq (Brazil) and CHS
was financed by CAPES (Brazil). We also thank an anonymous referee for calling our attention to Ref.~\cite{kiessling}.

\end{document}